\documentclass[trackchanges]{aastex701}

\usepackage{booktabs} 
\usepackage{multirow} 
\usepackage{makecell}
\usepackage{longtable}
\usepackage{booktabs}
\usepackage{makecell}
\usepackage{array}
\usepackage{hyperref}

\begin{document}

\title{Statistical Characterization of CME Interaction Responsible for Successive Type II Solar Radio Bursts}

\shorttitle{Successive Type II Solar Radio Bursts}

\author[orcid=0000-0001-5742-9033,gname=Anshu, sname='Kumari']{Anshu Kumari}
\affiliation{Udaipur Solar Observatory, Physical Research Laboratory, Dewali, Badi Road, Udaipur - 313001, Rajasthan, India}
\email[show]{anshu@prl.res.in}

\author[orcid=0009-0001-7689-0084, sname='Paliwal']{Divya Paliwal}
\affiliation{Udaipur Solar Observatory, Physical Research Laboratory, Dewali, Badi Road, Udaipur - 313001, Rajasthan, India}
\affiliation{Indian Institute of Technology, Gandhinagar, Gujarat-382355, India}
\email{divya@prl.res.in}  

\author[sname=Jaydeep]{Jaydeep Kandekar}
\affiliation{Department of Physics, Ahmednagar College, Station Road, Ahilyanagar, 414001 Maharashtra, India }
\email{citizenofbharat@outlook.com}

\begin{abstract}
Successive type II solar radio bursts are often considered as diagnostics for the evolution of CME-driven shocks and their interaction with the structured solar corona. However, the physical mechanisms driving these successive type II emission are not well understood, with prior studies mostly limited to single event. In this work, we provide the first statistical analysis of successive type II radio bursts over Solar Cycles 23, 24 and the ascending phase of Solar Cycle 25 (1997--2025). We have constructed a catalogue of 73 independent successive type II events using radio dynamic spectra and SOHO/LASCO coronagraph observations. The events were divided into three observational classes, based on the inferred physical interaction mechanisms: CME–Streamer interaction, CME–CME interaction and Single CME. Statistical analyses were performed on the 56 events with clear physical classification.  CME--streamer interactions constitute the largest class among the physically classified events, highlighting the potential importance of the ambient coronal environment in shaping successive type II
radio emission. Comparative statistical analyses indicate that the CME linear speed has marginal differences among the physical classes, but the CME angular width is significantly different between the CME--CME and CME--Streamer populations. Other factors, such as shock evolution and the ambient coronal environment, are important for the production of successive type II radio bursts. The catalogue presented here provides a statistical framework for the investigation of CME-driven shock evolution.

\end{abstract}


\section{Introduction} 
\label{sec:section1}

Type II solar radio bursts are one of the most important observational signatures of magnetohydrodynamic (MHD) shock waves propagating in the solar corona and interplanetary medium \citep{Wild1963}. These slowly drifting radio emissions are generally interpreted as the plasma radiation produced by electrons accelerated in coronal or interplanetary shocks. Since their discovery, type II bursts have been useful diagnostics of shock formation, particle acceleration, and large-scale evolution of solar eruptive events \citep{Dulk1985}. Based on observational and theoretical studies, most of the coronal type II radio bursts are found to be closely associated with coronal mass ejections (CMEs), while in a few events flare-generated blast waves may also contribute \citep{Kumari2023b}. A CME propagating through the structured solar corona can drive a shock when its propagation speed exceeds the local fast-mode magnetosonic speed \citep{2025A&A...703A.271Z}. The efficiency of electron acceleration and thus the radio emission depends on the intrinsic properties of the CME but also on the ambient coronal environment such as the magnetic field configuration, plasma density and Alfv{\'e}n speed \citep{Vrsnak2002}. As, solar corona is highly structured, having streamers, coronal holes, active region loops, and debris of past eruptions \citep{Giordano2008}, CME-driven shocks do not propagate in a uniform medium. The interaction with these large scale coronal structures may largely modify the shock strength and geometry, leading to a temporal variation of the radio emission. In particular, encounters with dense coronal streamers may locally reduce the Alfv\'{e}n speed, thus improving the shock and increasing the acceleration of electrons \citep{warmuth2005model}. Similarly, the interaction between consecutive CMEs can modify the upstream plasma conditions, leading to the formation of new shocks and multiple radio emission episodes \citep{2025NatSR..1534279V}.

Successive type II radio bursts are of particular interest among the various manifestations of shock-associated radio emission. As opposed to isolated type II bursts, successive events exhibit two or more distinct type II episodes associated with the same eruptive event. The physical origin of these multiple episodes remains a question. Previous studies have proposed many possible mechanisms, including interactions of successive CMEs \citep{2025NatSR..1534279V}, encounters of CME-driven shocks with coronal streamers \citep{Cho2011A}, and repeated shock formation during the evolution of a single CME \citep{Kumari2023b}. However, the relative importance of these mechanisms has not been statistically established yet. Most previous work on successive type II radio bursts has concentrated on case studies or small samples of events. Such studies have given useful insights into the physics of particular events, but have not addressed the question of whether different physical mechanisms are associated with systematically different CME properties. To date, a detailed statistical comparison of the kinematic and morphological properties of CMEs associated with different classes of successive type II bursts has not been performed. In this work we present a statistical study of 73 consecutive type II radio burst events for Solar Cycles (SC) 23, 24 and rise phase of SC25. The Large Angle and Spectrometric Coronagraph \citep[LASCO C2/C3;][]{Brueckner1995} on board Solar and Heliospheric Observatory \citep[SOHO;][]{do95} and the cor1/cor2 coronagraphs of the Sun–Earth Connection Coronal and Heliospheric Investigation \citep[SECCHI;][]{Howard2008} instrument suite on board STEREO--A catalogue observations were used to identify the most probable physical mechanism responsible for the successive radio emission in each event. The analysis led to the events being classified into three physically motivated categories: isolated CME events, CME–streamer interaction events, and CME–CME interaction events.

The main aim of this study is to investigate whether these different interaction mechanisms show statistically different CME properties. In particular, we explore whether the CME linear speed and the angular width are significantly different among the three classes and we compare their relative capability to discriminate among the proposed physical mechanisms. To make a strong inference, non-parametric statistical tests including the Kruskal-Wallis test, Dunn’s multiple-comparison procedure and Cliff’s delta effect size analysis are applied.
The rest of the paper is organised as follows: Section-\ref{sec:section2} presents the event selection, observational data sets, classification methodology and the statistical techniques. The statistical analysis of the properties of CMEs associated with successive type II radio bursts is presented in Section-\ref{sec:section3}. In Section-\ref{sec:section4} we discuss the physical implications of the results and summarise the principal conclusions in Section-\ref{sec:section5}.

\section{Data and Event Selection}
\label{sec:section2}

\subsection{Event Selection}
\label{sec:section2.1}

The present study is based on a catalogue of successive type II solar radio burst events identified during SC23, SC24 and the rise phase of SC25. The initial event list was compiled from published type II burst catalogues together with radio dynamic spectra obtained from ground-based radio observatories \footnote{\url{https://www.e-callisto.org/}},
\footnote{\url{https://secchirh.obspm.fr/spip.php?article11}} \citep{Morosan2021, Kumari2023b}. 
Successive type II radio burst events were identified from the complete type II burst catalogue by requiring that the onset of the second type II burst occurred within one hour of the termination of the first type II burst. This criterion ensures temporal continuity between the two bursts while minimizing the inclusion of unrelated solar eruptive events. Each candidate pair satisfying this condition was subsequently examined using radio dynamic spectra and SOHO/LASCO coronagraph observations to determine the associated CME configuration and physical interaction mechanism.
Each candidate event was visually inspected to confirm the presence of two or more distinct type II emission episodes associated with a single solar eruptive event. The solar eruption associated with each radio event was identified from the observations of the SOHO/LASCO coronagraphs and STEREO/SECCHI-cor1/cor2. 
Events with ambiguous interaction classifications, insufficient observational coverage, or incomplete CME parameters were excluded. In contrast to previous studies that primarily concentrated on individual events, the present work aims to investigate the statistical properties of successive type II bursts and to determine whether different physical mechanisms are associated with systematically different CME characteristics. The final catalogue contains 137 CMEs associated with the 73 radio events, indicating that several successive type II bursts involve multiple eruptions. 

\subsection{Data Sources}
\label{sec:section2.2}

We used observations from multi-wavelength instruments and multiple data archives to probe the physical origin of successive type II radio bursts. The basic CME parameters such as first appearance time, linear speed, angular width, and central position angle were obtained from the SOHO/LASCO CDAW CME catalogue \citep{Yashiro2004, Gopalswamy2009cdaw}. These measurements provided the basis for the statistical analysis presented in this work. The occurrence and the temporal evolution of successive type II radio bursts identified from radio dynamic spectra were taken from Space Weather Prediction Centre (SWPC)\footnote{\url{https://www.spaceweather.gov/products/solar-and-geophysical-event-reports}} maintained by the National Oceanic and Atmospheric Administration (NOAA). We also used several ground-based open data radio spectrographs to confirm the successive radio type II bursts. These observations helped us to identify individual type II events associated with a specific eruptive event. Together, coronagraph observations and radio dynamic spectra allowed to obtain complementary information on the eruptive event and the surrounding coronal environment which give the basis for a physically motivated classification of successive type II radio burst events. Table \ref{tab:table1} lists all the data summaries used in the present study. 

\begin{table}[ht]
\centering
\caption{Summary of the successive type II radio burst event catalogue used in this study.}
\label{tab:table1}
\begin{tabular}{lc}
\toprule
\textbf{Parameter} & \textbf{Value} \\
\midrule
Total successive type II events & 73 \\
Study period & 1997--2025 \\
Primary CME parameters & Linear speed, Angular width \\
Event categories & CME--Streamer, CME--CME, Single CME, Unclear, Not Visible \\
Categories included in statistical analysis & CME--Streamer, CME--CME, Single CME \\
Excluded categories & Unclear, Not Visible \\
Statistical methods & Kruskal--Wallis test, Dunn's post hoc test, Cliff's $\Delta$, Logistic regression \\
\bottomrule
\end{tabular}
\end{table}

\subsection{Classification of Successive Type II Events}
\label{sec:section2.3}

Each event in the catalogue was individually examined using the SOHO/LASCO coronagraph observations and radio dynamic spectra. The associated CME evolution was inspected manually to identify the most probable physical mechanism responsible for the occurrence of successive type II radio bursts. Based on the observational evidence, the events were classified into the following categories: i) Single CME, successive type II bursts associated with a single CME, without observational evidence for interaction with either another CME or a coronal streamer; ii) CME--Streamer Interaction, events in which the CME-driven shock encounters a coronal streamer during its outward propagation, streamer deflections, compression, or clear interactions observed in coronagraph images were used as indicators of this class; iii) CME--CME Interaction, events in which two or more CMEs interact during their propagation through the corona, the classification was based on temporal proximity, overlapping propagation directions, and observational evidence for interaction in LASCO images; iv) Unclear, events for which the available observations were insufficient to determine the interaction mechanism with confidence; v) Not Visible, events for which no reliable CME or interaction signature could be identified because of limited or unavailable observational evidence. The latter two categories were retained in the catalogue but excluded from the comparative statistical analysis.

\begin{figure}[ht!]
\centering
\includegraphics[width=0.5\textwidth]{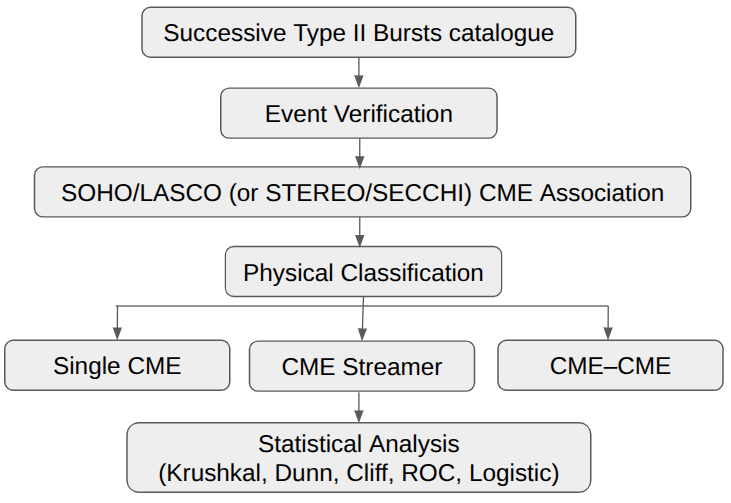}
\caption{
Flowchart of selection and classification procedure adopted in the present study. The event catalogue had 73 successive type II radio burst events initially.
}
\label{fig:figure1}
\end{figure}

\subsection{Statistical Methodology}
\label{sec:section2.4}

As the aim of this work is to study if the properties of CMEs are significantly different for the three proposed interaction mechanisms. The entire analysis was performed using non-parametric statistical techniques because of the non-Gaussian distributions of the CME parameters and the different sample sizes of the event classes. Differences between the three categories were initially tested by the Kruskal--Wallis rank-sum test \citep{doi:https://doi.org/10.1002/9780470479216.corpsy0491}. When differences were statistically significant, pairwise comparisons were made using Dunn's post-hoc test with Holm correction for multiple testing \citep{dunn1961multiple}. Statistical significance indicates whether an observed difference is unlikely to have occurred by chance, but it does not measure the magnitude of that difference. Thus, a non-parametric effect size measure, Cliff’s delta, was calculated for each pairwise comparison \citep{cliff1996answering}. Conventional thresholds for negligible, small, medium, and large effects were used to interpret the resulting effect sizes. Finally, the predictive power of CME linear speed and angular width to discriminate the different interaction mechanisms was evaluated through receiver operating characteristic (ROC) analysis \citep{gonen2006receiver} and logistic regression \citep{carter2016roc}. These analyses provide complementary information on the relative importance of the investigated CME parameters. Figure \ref{fig:figure1} shows the statistical flowchart of the current work carried out. 

\section{Results}
\label{sec:section3}

As we plan to determine whether the physical mechanisms responsible for successive type II radio bursts exhibit statistically distinguishable CME properties, here we first present an overview of the event sample before examining the distributions of CME linear speed and angular width for the different interaction classes. Finally, we compare the relative importance of these CME parameters and investigate their predictive capability using a statistical classification model.

\begin{figure}[!t]
\centering
\includegraphics[width=0.72\textwidth]{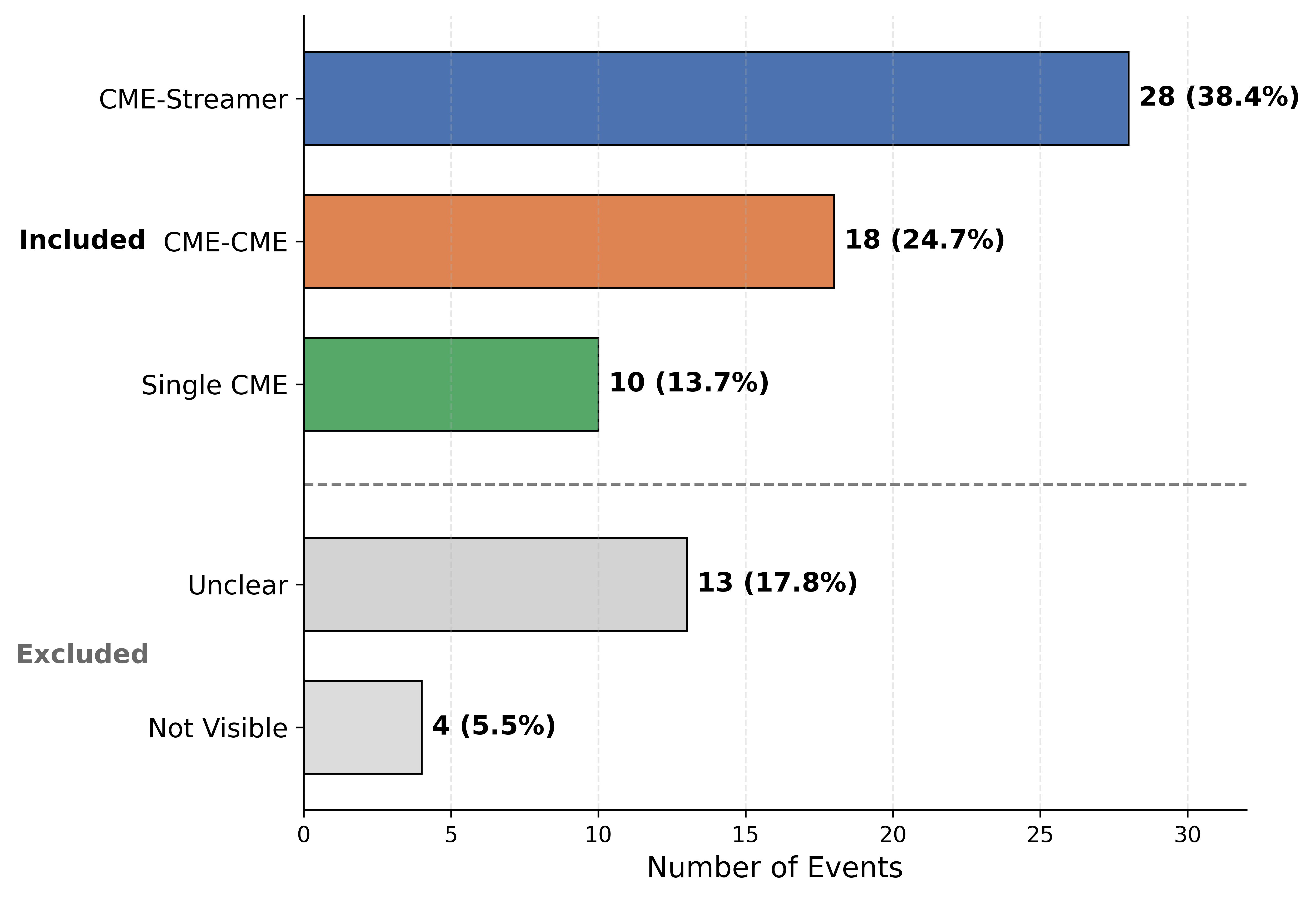}
\caption{
Distribution of the 73 successive type II radio burst events according to the inferred interaction mechanism. The majority of events are associated with CME--streamer interactions, followed by CME--CME interactions and single CME events. Events classified as \emph{Unclear} or \emph{Not Visible} were excluded from the subsequent statistical analyses.
}
\label{fig:figure2}
\end{figure}

\subsection{Overview of the Event Sample}
\label{sec:section3.1}

The final catalogue comprises 73 unique successive type II radio burst events identified during Solar Cycles 23, 24, and the rising phase of Solar Cycle 25. The methodology adopted in this study is summarized in Figure~\ref{fig:figure1}. Following the procedure described in Section-\ref{sec:section2.3}, each event was classified into one of five observational categories according to the inferred interaction mechanism. The resulting distribution of events is presented in Figure~\ref{fig:figure2}. Table \ref{tab:table2} contains the distribution of CME interaction mechanisms associated with successive type II radio bursts, with the not visible/unclear events omitted from further analysis.

\begin{table}[ht]
\centering
\caption{Distribution of CME interaction mechanisms associated with successive type II radio bursts. Statistical analyses were performed only for events with unambiguous classifications.}
\label{tab:table2}
\begin{tabular}{lcc}
\toprule
Interaction Mechanism & Number of Events & Percentage (\%) \\
\midrule
CME--Streamer & 28 & 50.0 \\
CME--CME & 18 & 32.1 \\
Single CME & 10 & 17.9 \\
Not Visible & 4 & --- \\
Unclear & 13 & --- \\
\bottomrule
\end{tabular}
\end{table}

\subsection{Physical Classification of Successive Type II Bursts}
\label{sec:section3.2}

\begin{figure*}[htbp]
\centering
\includegraphics[width=0.95\textwidth]{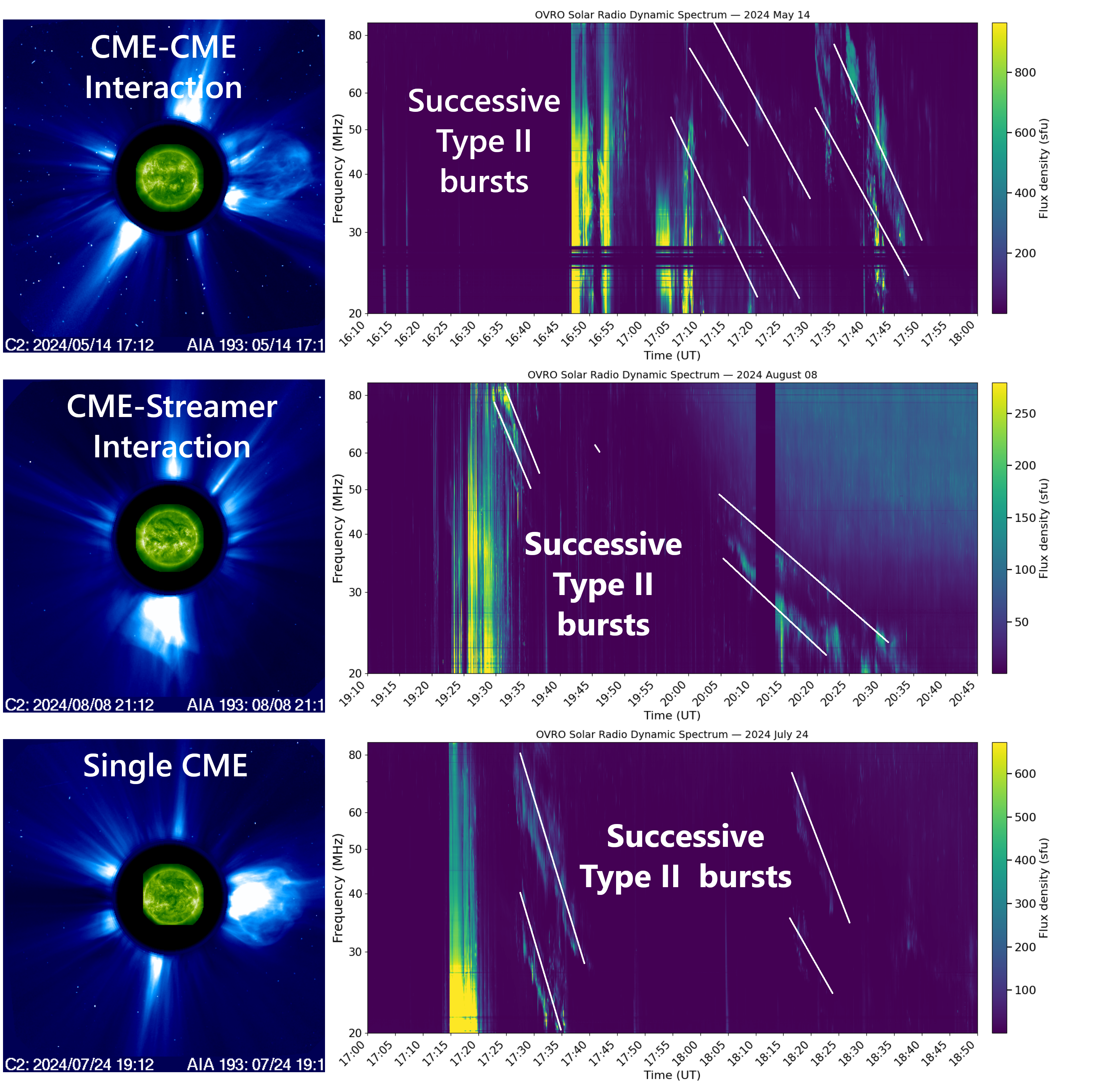}
\caption{
Examples of successive type II radio bursts associated with different CME configurations. The right panels show the radio dynamic spectra, while the left panels present the corresponding SOHO/LASCO C2 white-light coronagraph observations. Top panels correspond to the CME--CME interaction event of 14 May 2024; middle panels correspond to the CME--Streamer interaction event on 8 Aug 2024 and bottom panels correspond to the Single CME event of  24 Jul 2024. The dynamic spectra show the successive type II emission episodes, while the coronagraph observations illustrate the CME morphology and the surrounding coronal environment. 
}
\label{fig:representative_events}
\end{figure*}

The 73 events of consecutive type II radio bursts were classified based on the deduced physical mechanism of the repeated radio emission (see Appendix). For each event we assigned one of five observational categories according to the observational criteria described in Section-\ref{sec:section2.3}: Single CME, CME–Streamer interaction, CME–CME interaction, Unclear, or Not Visible. The \emph{Unclear} category refers to events for which the available observations were not sufficient to discern the interaction mechanism with confidence, while the \emph{Not Visible} category refers to events for which the associated CME or interaction process could not be reliably identified due to limited observational evidence. The last two categories were included in the catalogue for completeness, but were excluded from the comparative statistical analyses.

The distribution of the classified events is summarised in Figure~\ref{fig:figure2}. Of the 73 events, 28 (38.4\%) were CME--Streamer interaction events, 18 (24.7\%) were CME--CME interaction events and 10 (13.7\%) were Single CME events. Thirteen events (17.8\%) were classified as \emph{Unclear} and four events (5.5\%) were classified as \emph{Not Visible}. The dataset for the statistical analyses that follow consists of the physically classified three categories (CME--Streamer, CME--CME, and Single CME). In the following sections we compare the CME kinematic properties of these three classes to investigate whether different physical mechanisms responsible for successive type II radio bursts are associated with systematically different CME characteristics.

\subsection{Representative Events and Physical Interpretation
of Successive Type II Bursts}
\label{sec:representative_events}

To complement the statistical analysis and investigate the physical processes responsible for successive type II radio bursts, we selected three representative events corresponding to the Single CME, CME--CME interaction, and CME--Streamer interaction categories. For each event, the radio dynamic spectra were examined together with the corresponding
SOHO/LASCO C2 white-light coronagraph observations. Figure~\ref{fig:representative_events} presents these
examples, illustrating the different coronal configurations
associated with successive type II emission. \footnote{The spectral data is obtained from OVRO-LWA solar observations; hence, those are limited to 25-85 MHz here in the plots. However, the list provided int he appedix is extracted from SWPC event list, hence it ranges from 25-180 MHz \url{https://www.ovsa.njit.edu/lwa/}.}

\subsubsection{Single CME Event}
The event of 2024 July 24 represents the Single CME category (Figure~\ref{fig:representative_events}, panels e and f). The OVRO-LWA dynamic spectrum, covering approximately 17:00--18:50 UT, shows two temporally separated, downward-drifting radio emission features identified as successive type II bursts. The first type II burst was observed from 17:26 to 17:37 UT over 25--152 MHz, followed
by a second type II burst from 18:16 to 18:24 UT over 25--72 MHz. The corresponding SOHO/LASCO C2 image obtained at 19:12 UT shows an extended white-light CME structure predominantly toward the western side of the coronagraph field of view. 

The occurrence of successive type II bursts in association with a single CME suggests that repeated shock-related radio emission does not necessarily require CME--CME interaction. Instead, the two emission episodes may arise from the temporal evolution of the same CME-driven shock or from spatially
separated regions of its expanding front. Changes in the local plasma density, Alfv\'en speed, and shock geometry may alter the efficiency of electron acceleration, resulting in distinct episodes of type II
emission. The two drifting features observed in the dynamic spectrum are therefore consistent with the possibility of repeated electron acceleration at different stages or locations of the evolving CME-driven shock. However, without spatially resolved radio observations, the precise locations of
the two radio-emitting regions cannot be established. 

\subsubsection{CME--CME Interaction Event}
The event of 2024 May 14 illustrates the
CME--CME interaction category (Figure~\ref{fig:representative_events}.
The OVRO-LWA dynamic spectrum covers approximately 16:10--18:00 UT and reveals multiple downward-drifting emission features, including two successive type II episodes. The first type II burst occurred between
16:50 and 17:15 UT over 25--52 MHz, whereas the second burst was detected
between 17:30 and 18:00 UT over 25--180 MHz. The SOHO/LASCO C2 image obtained at 17:12 UT reveals a complex white-light coronal structure with extended emission toward the western side of the field of view.
The first associated halo CME was observed at
17:00 UT with a linear speed of 2018 km s$^{-1}$.
The second halo CME appeared at 17:48 UT with a linear speed of 1407 km s$^{-1}$.

The CME--CME interaction scenario provides a possible explanation for the successive radio emission. When a later CME propagates
into the disturbed coronal environment created by a preceding eruption, the
upstream plasma density, magnetic-field configuration, and local Alfv\'en speed may be modified. These changes can influence the shock strength, geometry, and electron acceleration efficiency, potentially producing renewed type II emission. The presence of successive drifting
features in the radio spectrum is consistent with repeated episodes of shock-associated electron acceleration.

\subsubsection{CME--Streamer Interaction Event}
The event of 2024 August 8 represents the CME--Streamer interaction category (Figure~\ref{fig:representative_events}. The OVRO-LWA dynamic spectrum covers approximately 19:10--20:45 UT and shows two distinct periods of downward-drifting radio emission. The first type II burst was observed
between 19:29 and 19:35 UT over 62--180 MHz, followed by the second
burst between 20:05 and 20:15 UT over 25--49 MHz. The corresponding SOHO/LASCO C2 image obtained at 21:12 UT shows a bright CME structure extending predominantly toward the southern corona, together with surrounding
large-scale coronal structures. The associated CME has a catalogued
linear speed of 789 km s$^{-1}$.

In the CME--Streamer interaction scenario, the expanding CME-driven
shock may encounter a relatively dense coronal streamer, modifying the local shock properties. Variations in the upstream plasma density and Alfv\'en speed may change the shock Mach number and the conditions for electron
acceleration. Consequently, the interaction between a CME-driven shock and a streamer may produce enhanced or renewed type II radio emission. The two downward-drifting features observed in the dynamic spectrum are consistent with distinct episodes of shock-associated radio emission. However, the
precise association of the second type II burst with a CME--Streamer interaction requires confirmation from the temporal evolution of the coronagraph observations.

\subsection{CME Linear Speed}
\label{sec:section3.3}

\begin{figure}[!t]
\centering
\includegraphics[width=0.48\textwidth]{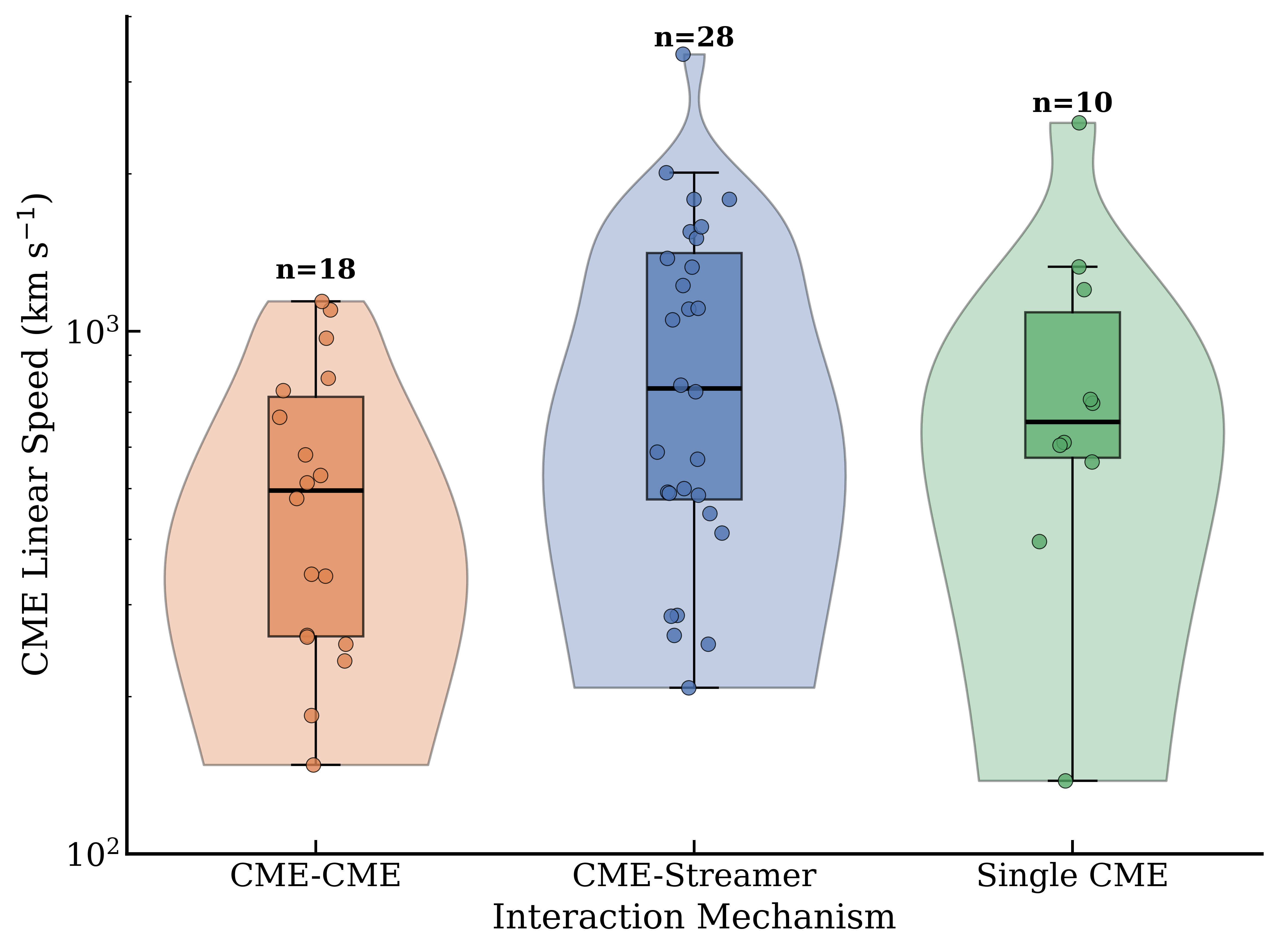}
\includegraphics[width=0.48\textwidth]{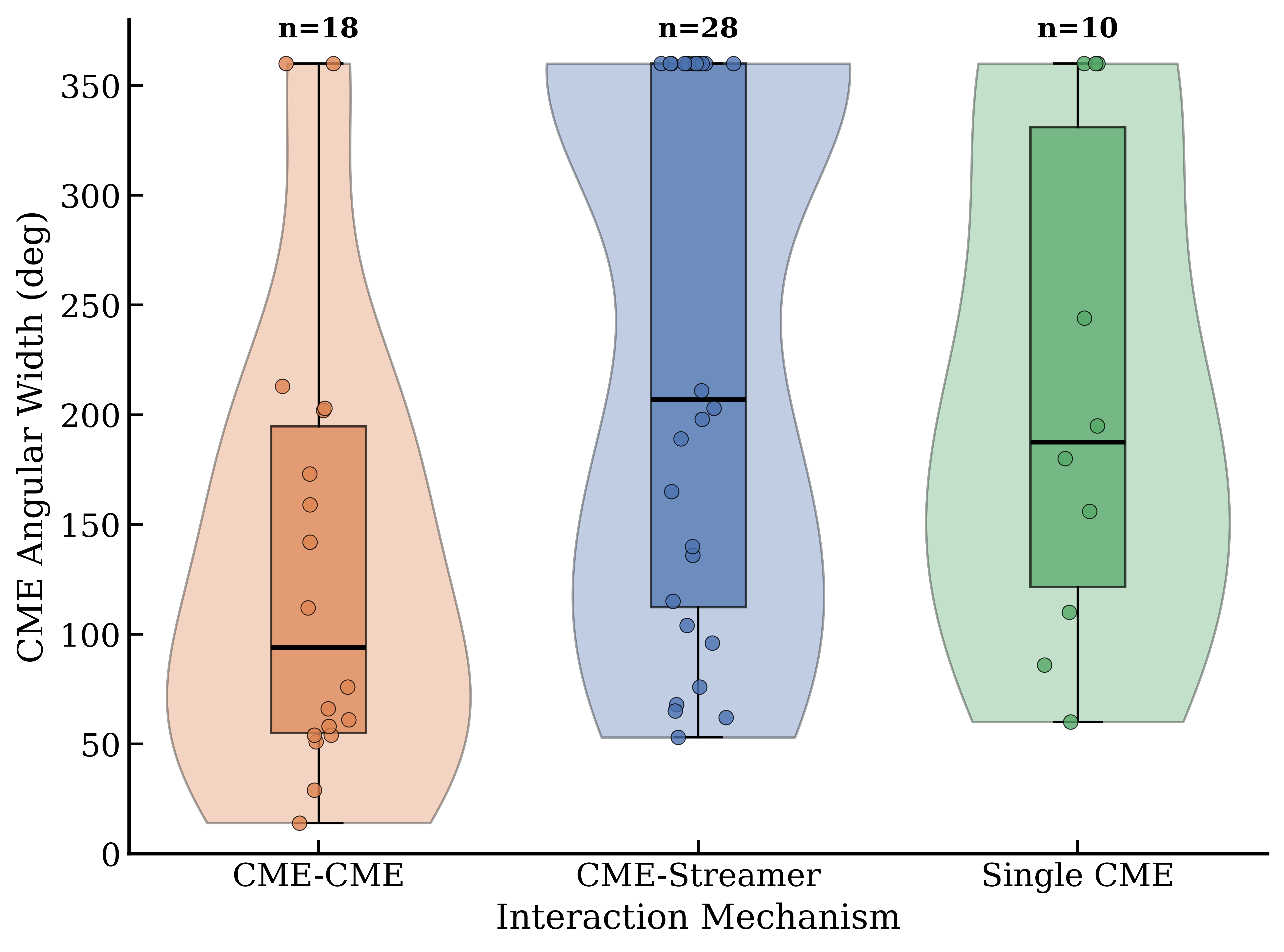}
\caption{
Left panel: Distribution of CME linear speeds for the three physically classified categories of successive type II radio burst events: CME--Streamer interaction, CME--CME interaction, and Single CME events. Violin plots represent the kernel density estimates of the speed distributions, while the embedded boxplots indicate the median and interquartile range. Individual points correspond to the measured CME linear speeds for each event. Right panel: Distribution of CME angular widths for the three interaction mechanisms associated with successive type II radio bursts. Violin plots represent the kernel density estimate of each distribution, while the embedded boxplots indicate the median and interquartile range. Individual symbols correspond to individual CME events. CME--Streamer and Single CME events are generally associated with wider CMEs than CME--CME interaction events, consistent with the non-parametric statistical analyses presented in Section~3.
}
\label{fig:figure3}
\end{figure}

One of the main controlling parameters for the development and progression of CME-driven shocks is the linear speed of a coronal mass ejection. More energetic shocks are expected to be produced by faster CMEs, which are more efficient at accelerating electrons and therefore are expected to enhance the type II radio emission. To examine whether the physical mechanisms that produce the successive type II radio bursts are associated with different CME kinematics, we compared the distributions of the CME linear speeds of the three physically classified event categories. The distribution of CME linear speeds for CME-CME interaction events, CME-Streamer interaction events and Single CME events are shown in Figure~\ref{fig:figure3}a. The median CME speed is highest for CME--Streamer interaction events (777 km s$^{-1}$), followed by Single CME events (670 km s$^{-1}$), and lowest for CME--CME interaction events (496 km s$^{-1}$). The corresponding mean CME speeds are 986 km~s$^{-1}$, 881 km~s$^{-1}$ and 533 km~s$^{-1}$ for the CME--Streamer, Single CME and CME--CME categories respectively.

The differences were tested for statistical significance using the non-parametric Kruskal--Wallis test that yielded the test statistic $H = 5.74$ and the corresponding probability $p = 0.056$. The differences seen are near the traditional 5\% significance level, but do not provide enough evidence to reject the null hypothesis that the three samples are from the same parent population. Dunn’s post hoc test with Holm correction was then performed for pairwise comparisons. The difference between CME--CME interaction events and CME--Streamer interaction events was close to statistical significance ($p = 0.059$), but the difference between CME--CME interaction events and Single CME events ($p = 0.23$) and the difference between CME--Streamer interaction events and Single CME events ($p = 0.83$) were not statistically significant. Cliff's $\Delta$ was calculated for each pair of categories to measure how different the observed differences were. The comparison of CME--CME interaction events to CME--Streamer interaction events ($\Delta = -0.41$) and CME--CME interaction events to Single CME events ($\Delta = -0.37$) had a medium effect size while the comparison of CME--Streamer interaction events to Single CME events had a negligible effect size ($\Delta = 0.05$). The results show that CME--Streamer and Single CME events tend to have higher CME speeds than CME--CME interaction events, but the large overlap between the distributions limits the discriminatory power of CME speed alone.

\begin{table*}[ht]
\centering
\caption{Summary statistics of CME linear speed and angular width for the three physically classified interaction mechanisms. The values shown correspond to the 56 independent events used in the statistical analyses.}
\label{tab:statistics}
\begin{tabular}{lccccccc}
\toprule
Category &
$N$ &
Median Speed &
Q1--Q3 Speed &
Mean Speed &
Median Width &
Q1--Q3 Width &
Mean Width \\
&
&
(km s$^{-1}$) &
(km s$^{-1}$) &
(km s$^{-1}$) &
($^\circ$) &
($^\circ$) &
($^\circ$)\\
\midrule
CME--CME &
18 &
496 &
260--748 &
533 &
94 &
55--194 &
132 \\

CME--Streamer &
28 &
777 &
476--1410 &
986 &
207 &
112--360 &
234 \\

Single CME &
10 &
670 &
572--1086 &
881 &
187 &
121--331 &
211 \\

\bottomrule
\end{tabular}
\end{table*}

\subsection{CME Angular Width}
\label{sec:section3.4}

The angular width of a coronal mass ejection is a measure of its spatial extent in the corona and is often used as an indicator of the scale and geometry of the eruptive event. Large CMEs typically have larger shock surfaces and are more likely to interact with surrounding coronal structures, e.g. streamers or preceding CMEs. We therefore investigate if the angular widths of CMEs were systematically different among the three physical classes of successive type II radio burst events. The distributions of CME angular widths for CME--CME interaction events, CME--Streamer interaction events, and Single CME events are shown in figure~\ref{fig:figure3}b. CME–Streamer interaction events have the largest median angular width (207$^{\circ}$) followed by Single CME events (187$^{\circ}$), while CME–CME interaction events have the smallest median width (94$^{\circ}$). The corresponding average angular widths are 234$^{\circ}$, 211$^{\circ}$, and 132$^{\circ}$ for the CME--Streamer, Single CME, and CME--CME categories, respectively.

We tested significance of the observed differences using a Kruskal-Wallis rank-sum test. This analysis yields a test statistic of $H = 7.91$, with an associated probability of $p = 0.019$. We find that the distributions of angular widths of the three physical classes are significantly different. Dunn’s post hoc test with Holm correction was used to perform pairwise comparisons to determine which groups differed. The CME–CME interaction events are statistically different from the CME–Streamer interaction events ($p = 0.017$), while the CME–CME interaction events are not statistically different from the Single CME events ($p = 0.163$), and the CME–Streamer interaction events are not statistically different from the Single CME events ($p = 0.69$). The magnitude of the differences was measured with Cliff's $\Delta$. Medium effect sizes are found for the comparisons between CME--CME interaction events and CME--Streamer interaction events ($\Delta = -0.47$) and between CME--CME interaction events and Single CME events ($\Delta = -0.42$), while the comparison between CME--Streamer interaction events and Single CME events gives a negligible effect size ($\Delta = 0.10$). These results suggest that CME angular width provides better separation between the physical classes than CME linear speed, particularly in distinguishing CME--Streamer interaction events from CME--CME interaction events.

\begin{figure}[!t]
\centering
\includegraphics[width=0.48\textwidth]{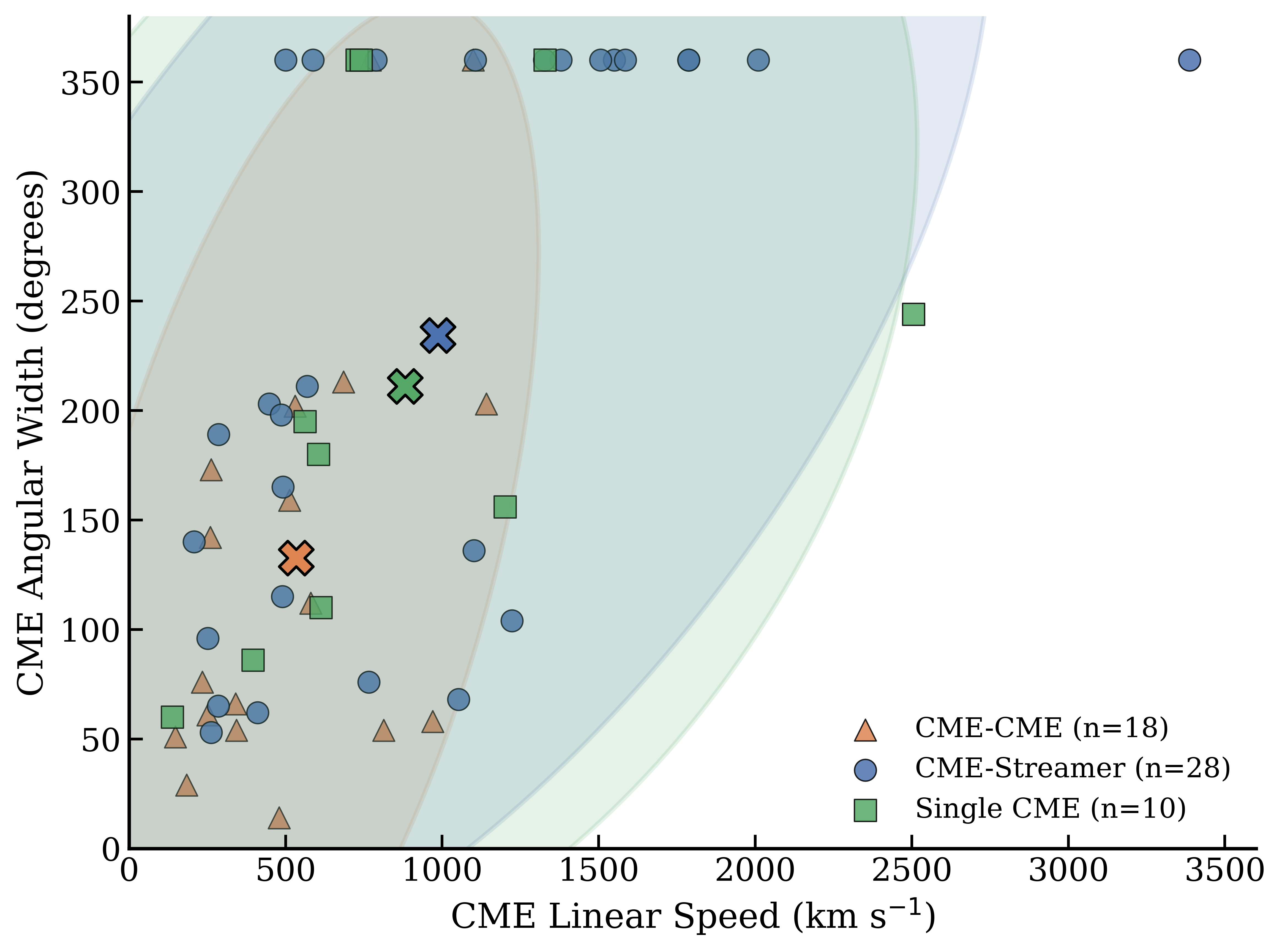}
\caption{Relation between CME linear speed and angular width for the three physically classified interaction mechanisms associated with successive type II radio bursts. Orange triangles denote CME--CME interaction events, blue circles represent CME--Streamer interaction events, and green squares correspond to Single CME events. The colored X symbols indicate the centroid (mean CME speed and mean angular width) of each class, while the shaded ellipses represent the approximate 95\% confidence regions estimated from the sample covariance. CME--CME interaction events are preferentially located toward lower CME speeds and narrower angular widths, whereas CME--Streamer and Single CME events occupy broader and largely overlapping regions of the parameter space. The substantial overlap between the latter two populations illustrates that CME linear speed and angular width alone are insufficient to uniquely distinguish the physical mechanisms responsible for successive type II radio bursts.
}
\label{fig:figure4}
\end{figure}

\subsection{Comparative Analysis of CME Speed and Angular Width}
\label{sec:section3.5}

The preceding sections discussed CME linear speed and angular width separately, but the two parameters might contain complementary information when considered together. Figure~\ref{fig:figure4} shows the joint distribution of the 56 physically classified successive type II radio burst events in the two dimensional parameter space of CME linear speed and angular width. The three classes of interactions show different trends. CME--Streamer interaction events are preferentially quoted in the region of relatively high CME speeds and large angular widths, consistent with energetic eruptions capable of driving shocks over extended regions of the corona and interacting with large-scale coronal streamers. Conversely, CME–CME interaction events tend to be more biased toward lower CME speeds and narrower angular widths. The single CME events largely overlap with the CME-Streamer population with similar ranges of CME speed and angular width.

However, there is significant overlap between the three categories, despite these systematic trends, which implies that neither the CME speed nor the angular width alone is enough to provide complete separation of the physical mechanisms responsible for successive type II radio bursts. However, the clustering found suggests that the combination of the two parameters contains useful discriminating information. The larger differences in angular width relative to the CME speed are consistent with the statistical analyses of Sections~\ref{sec:section3.3} and \ref{sec:section3.4} that only the statistical differences in the CME angular width among the interaction classes are statistically significant, while those in the CME speed are only marginally significant. The partial separation seen in Figure~\ref{fig:figure4} inspires the multivariate classification analysis discussed in the next section, where the combined predictive capacity of CME speed and angular width is quantitatively evaluated.

\subsection{Exploratory Classification Analysis}
\label{sec:section3.6}

The statistical analyses of the previous sections show systematic differences in the CME kinematic properties associated with the three physical classes of successive type II radio burst events. As an exploratory exercise, we have investigated if two easily available CME parameters, the linear speed and the angular width, contain enough information to automatically discriminate between these interaction mechanisms.

Based on the 56 physically classified events (18 CME–CME, 28 CME–Streamer, and 10 Single CME events), we constructed a multinomial logistic regression classifier with CME linear speed and angular width as predictors. Because of the small sample size and the resulting class imbalance, the classifier was evaluated by means of repeated stratified five-fold cross-validation to obtain a robust estimate of its predictive performance. The classifier reached a mean classification accuracy of $0.54 \pm 0.11$ with macro averaged precision, recall and F$_1$-score of $0.36 \pm 0.10$, $0.43 \pm 0.10$, and $0.38 \pm 0.09$, respectively. While these values are above what would be expected by chance for a three class problem, they suggest only moderate discriminatory ability. This is consistent with the large overlap of the CME--Streamer and Single CME populations in both the CME speed and angular width distributions (Figures~\ref{fig:figure4}, while the CME--CME events occupy a relatively distinct region of lower speeds and narrower angular widths.

It is important to point out that the present classification experiment is only intended as an exploratory assessment of the information content in CME speed and angular width. The number of events available for training is relatively small, especially for the Single CME class, so the classification metrics should not be taken as indicative. Rather they show that these two CME parameters alone are not enough to uniquely determine the physical mechanism behind successive type II radio bursts. The low predictive performance indicates that the generation of successive type II radio bursts depends on other physical parameters than the global CME kinematics. The interaction process and the related radio emission are likely to be affected by factors like CME acceleration, propagation direction, shock evolution, coronal magnetic topology, streamer geometry and the ambient coronal plasma environment. Thus, while CME linear speed and angular width are helpful first-order diagnostics, a more complete description of CME–corona interaction will require the inclusion of these additional physical parameters in future studies.

\begin{table*}[ht]
\centering
\caption{Summary of the statistical comparisons of CME linear speed and angular width among the three physical classes of successive type II radio burst events. Pairwise comparisons were performed using Dunn's post hoc test with Holm correction. Cliff's $\Delta$ is reported as the corresponding non-parametric effect size.}
\label{tab:statistical_tests}
\begin{tabular}{lcccc}
\toprule
Parameter &
Comparison &
Dunn's $p$-value &
Cliff's $\Delta$ &
Effect Size \\
\midrule

\multirow{3}{*}{Linear Speed}
& CME--CME vs CME--Streamer & 0.06 & $-0.41$ & Medium \\
& CME--CME vs Single CME    & 0.23 & $-0.37$ & Medium \\
& CME--Streamer vs Single CME & 0.83 & $0.05$ & Negligible \\

\midrule

\multirow{3}{*}{Angular Width}
& CME--CME vs CME--Streamer & 0.02 & $-0.47$ & Medium \\
& CME--CME vs Single CME    & 0.16 & $-0.42$ & Medium \\
& CME--Streamer vs Single CME & 0.69 & $0.10$ & Negligible \\

\bottomrule
\end{tabular}
\end{table*}

Table~\ref{tab:statistical_tests} summarizes the results of the pairwise statistical comparisons. Although both CME linear speed and angular width exhibit medium effect sizes between CME--CME interaction events and the other two physical classes, only CME angular width remains statistically significant after multiple-comparison correction. In contrast, negligible effect sizes and non-significant $p$-values are obtained for the comparison between CME--Streamer and Single CME events for both parameters. These results indicate that CME angular width provides a more robust discriminator of the physical mechanisms responsible for successive type II radio bursts than CME linear speed.

\section{Discussion}
\label{sec:section4}

\subsection{Physical Interpretation of CME--Streamer Interactions}
\label{sec:section4.1}

CME--streamer interactions constitute the largest group among the
physically classified successive type II radio burst events in our
catalogue, accounting for 28 of the 56 classified events (50\%). This observational result emphasises the possible role of the large-scale coronal environment in determining the evolution of CME-driven shocks and its associated radio emission. However, the high relative abundance of this class in our sample should not be interpreted as a direct measure of the probability of successive type II bursts by a CME-streamer interaction. This is because the present study does not include a control sample of all CME-streamer interactions without successive type II emission.

Coronal streamers are elongated structures with an enhanced plasma density and a particular magnetic-field configuration. Because they have long lifetimes they provide pre-existing structures that can be hit by the nose or flanks of an expanding CME and its associated shock. Such encounters can alter the local conditions of the shock propagation. Specifically, the local fast-mode magnetosonic speed is a function of the magnetic field and plasma density, and the shock strength is a function of the shock speed relative to the upstream magnetosonic speed. Thus, a streamer encountering a CME-driven shock may modify the local Mach number of the shock, its geometry, and the efficiency of electron acceleration.

Type II radio emission results from energetic electrons accelerated at shock fronts. The emission is usually understood as plasma emission at or near the local plasma frequency or its harmonic. The plasma frequency depends on the electron density, so propagation through a structured corona can also affect the observed frequency evolution of the emission. Therefore, changes in the local shock conditions and electron-acceleration efficiency during a streamer encounter may lead to renewed or enhanced radio emission, which may contribute to the appearance of distinct type II episodes. But the presence of successive bursts does not mean that a completely new shock is forming: the observations may also be changes in the radio-emitting regions of a shock that is evolving. This interpretation is consistent with the relatively large angular widths observed for the CME–streamer population. A wider CME and its associated shock can span a larger range of position angles and thus might impact an extended coronal structure over a larger part of the shock surface. This provides a plausible physical explanation for the larger median angular width of the CME--streamer events as compared to the CME--CME events. However, the CME angular width itself does not guarantee that a streamer encounter took place or that it gave rise to the successive radio emission. The classification in this study is based on the combined observational evidence from the radio dynamic spectra and the coronagraph images. The large spatial extent of streamer structures increases the probability of interaction with shocks, compared to the relatively infrequent occurrence of closely spaced CMEs that can interact in the low corona \citep{Cho2011A}. The high frequency of CME-streamer interaction events highlights the necessity to take into account the ambient coronal environment when interpreting subsequent type II radio bursts. The repeated type II emission seems to be strongly dependent on the coronal structures encountered by the propagating shock and not only on the properties of the erupting CME.

These results together suggest that the ambient corona is an important factor in the interpretation of successive type II radio bursts. The prevalence of CME--streamer events in the present catalogue warrants further investigation into the function of coronal density structures, magnetic topology and shock evolution in the production of repeated radio emission.

\subsection{CME Kinematics, Angular Width, and CME--CME Interactions}
\label{sec:section4.2}

The statistical analyses reveal that CME angular width is more discriminating than the CME linear speed. The speed distributions of the three interaction classes are only marginally different in a statistical sense, but the CME angular width is significantly different between the CME--CME and CME--Streamer populations. This implies that the physical mechanism responsible for successive type II radio bursts is more correlated with the spatial extent of a CME than with the bulk propagation speed. 

CME--CME interactions provide a unique way to modify the shock environment. If a later CME encounters the disturbed plasma and magnetic field from an earlier eruption, the conditions ahead of its shock can be very different from those in the undisturbed corona. The CMEs can interact and merge depending on their relative speeds, propagation directions and separation of the eruptions, and the associated shocks can evolve as they propagate through the resulting environment. These changes can impact the local shock strength and electron-acceleration conditions, and may generate separate episodes of type II emission. However, the mere existence of two CMEs is not sufficient to say that the interaction between them is responsible for the successive bursts. Temporal and morphological evidence of an interaction is also needed. The median angular width of CME--CME events in our sample is $94^\circ$, whereas for CME--streamer events it is $207^\circ$.
The difference in angular width between these two classes is statistically significant after the applied multiple-comparison correction, while the corresponding difference in speed is not.
One interpretation is that the CME--CME class comprises a larger fraction of events involving relatively narrow eruptions whose radio emission is affected by interaction between successive ejecta, rather than by the broad spatial encounter of a shock with a large-scale streamer. This interpretation is tentative, because the measured angular width is a projected coronagraph quantity, so it does not fully describe the three-dimensional geometry of the CME or its shock.

The Single CME category shows that a pair of consecutive type II radio bursts is not necessarily associated with interaction of two distinct CMEs.
A single shock driven by a CME may propagate through a structured corona where the plasma density, magnetic-field orientation and magnetosonic speed vary with position. Different parts of the shock may thus experience different conditions for electron acceleration and radio emission as the eruption develops. Based on the available observations, this may result in multiple identifiable radio episodes even if there is no interaction with another CME or a streamer. The Single CME classification must therefore be seen in observational terms, it represents the lack of a clearly identified external interaction in the available data, not evidence that the shock propagates through a uniform or interaction-free corona.

In general, larger CME corresponds to larger shock surface propagating through the corona. Therefore, during their evolution, wide CMEs are more probable to cross coronal streamers, to meet preceding ejecta, or to interact with multiple large-scale magnetic structures. These interactions increase the probability of repeated episodes of shock formation and electron acceleration to subsequently generate successive type II radio emission. In contrast, the CME speed mainly determines the strength of the propagating shock but does not uniquely determine the coronal environment through which the shock propagates. The large overlap in speed and angular width between the CME–streamer and Single CME populations is consistent with these interpretations. Different coronal environments may be associated with the same global CME properties and the presence of successive type II emission may depend on local shock conditions not described by the two global CME parameters considered here. Hence, while the angular width provides useful statistical information, neither the speed nor the width alone can identify the physical mechanism responsible for the successive type II radio bursts.

\subsection{Relationship between Type II Frequency Drift and CME Speed}
\label{sec:drift_cme_speed}

To investigate the relationship between the frequency evolution
of successive type II radio bursts and the kinematic properties
of their associated CMEs, we estimated the average frequency-drift
magnitude separately for the first and second type II bursts.
Using the frequency ranges and durations reported in
Table~\ref{tab:successive_typeII_catalogue}, the average
frequency-drift magnitude was approximated as

\begin{equation}
D = \left|\frac{\Delta f}{\Delta t}\right|
  = \frac{f_{\rm high}-f_{\rm low}}
         {t_{\rm end}-t_{\rm start}},
\label{eq:average_drift}
\end{equation}

where $f_{\rm high}$ and $f_{\rm low}$ represent the upper
and lower frequency limits of the observed emission,
respectively, and $t_{\rm start}$ and $t_{\rm end}$ denote
the burst onset and end times. This approximation assumes
that the emission drifts from higher to lower frequencies
over the reported duration. Consequently, $D$ represents
an estimated average frequency-drift magnitude rather than
a drift rate obtained by fitting an individual emission lane
in the dynamic spectrum.

We compared these estimates with the CME linear speeds
reported in the SOHO/LASCO CDAW catalogue. The analysis
was restricted to the 56 successive type II events classified
as CME--Streamer interactions, CME--CME interactions,
or Single CME events. The event classifications were adopted
from Table~\ref{tab:successive_typeII_catalogue}.
For consistency, the linear speed of the first associated CME
was used for each event, including CME--CME interactions.
Events with unavailable CME speeds or undefined frequency-drift
estimates were excluded from the corresponding correlation
analysis.

\begin{figure*}[htbp]
\centering
\includegraphics[width=0.88\textwidth]
{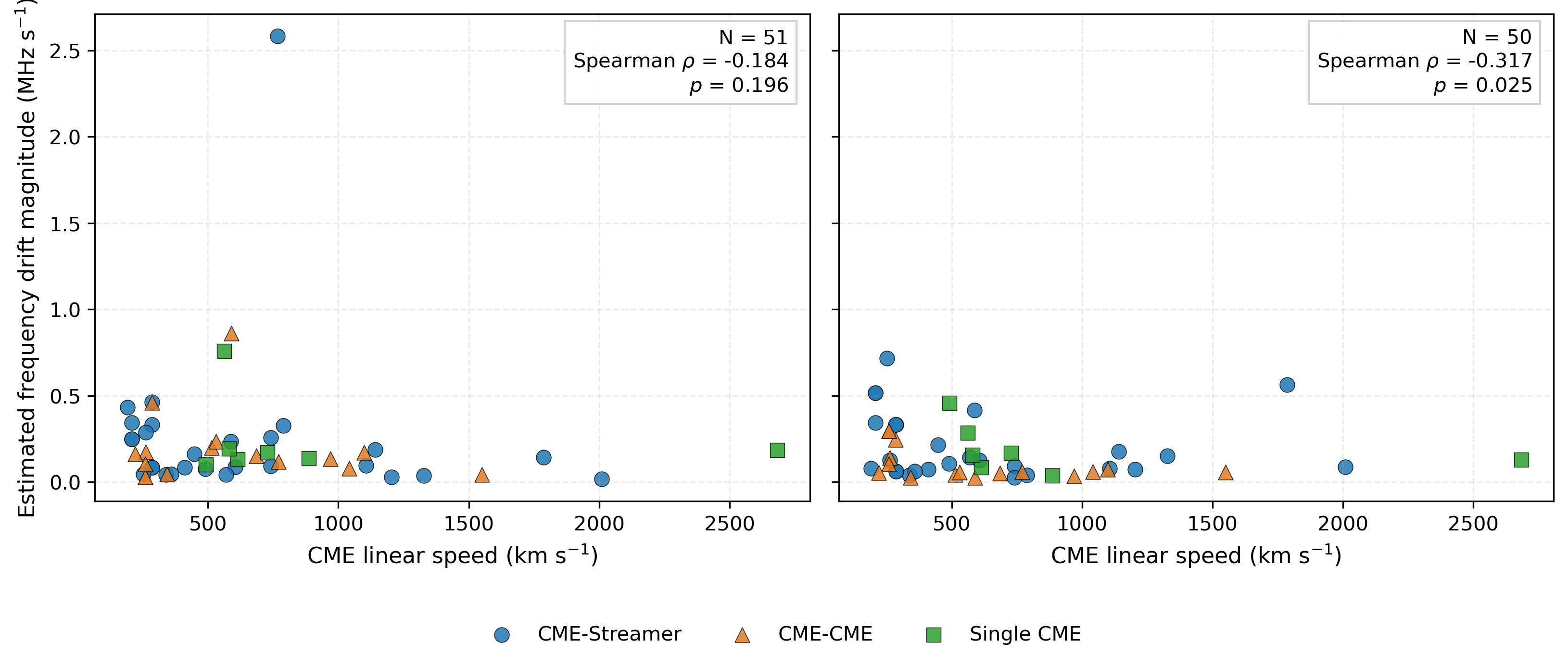}
\caption{
Relationship between the estimated average frequency-drift magnitude of successive type II radio bursts and the associated SOHO/LASCO CME linear speed. Panel (a) shows the first type II burst, while panel (b) shows the second type II burst.
Blue circles, orange triangles, and green squares represent
CME--Streamer interactions, CME--CME interactions,
and Single CME events, respectively. The Spearman rank correlation coefficient ($\rho$), corresponding two-sided $p$-value, and number of valid event pairs ($N$) are indicated in each panel. The first type II bursts exhibit a weak, statistically non-significant negative correlation
($\rho = -0.184$, $p = 0.196$, $N = 51$),
whereas the second type II bursts show a moderate
negative correlation ($\rho = -0.317$, $p = 0.025$, $N = 50$).
The frequency-drift magnitudes are estimated from
the catalogue frequency intervals and burst durations
rather than from direct spectral-lane fitting.
}
\label{fig:drift_speed_corr}
\end{figure*}

Figure~\ref{fig:drift_speed_corr} presents the relationship
between the estimated frequency-drift magnitude and the
associated CME linear speed for the first and second
type II bursts. We used Spearman's rank correlation
coefficient to quantify the monotonic association between
these parameters. For the first type II bursts, the analysis
includes 51 valid event pairs and yields a weak negative
correlation ($\rho = -0.184$, $p = 0.196$), indicating no
statistically significant monotonic relationship.
For the second type II bursts, 50 valid event pairs
yield a moderate negative correlation
($\rho = -0.317$, $p = 0.025$).
Although the latter correlation is nominally significant,
its interpretation requires caution because the frequency-drift
values are approximate and some radio-burst pairings may
share the same associated CME.

The lack of a positive correlation between the estimated
frequency-drift magnitude and CME linear speed suggests
that the observed spectral evolution cannot be explained
by CME propagation speed alone. In particular, the negative
correlation obtained for the second type II bursts should
not be interpreted as evidence that faster CMEs drive
slower shocks. The frequency drift of plasma emission
depends on both the propagation speed of the radio-emitting
shock and the spatial gradient of the ambient electron
density, which can be expressed as

\begin{equation}
\frac{df}{dt}
=
\frac{df}{dr}
\frac{dr}{dt},
\label{eq:drift_density}
\end{equation}

where $df/dr$ describes the frequency gradient associated
with the coronal density distribution and $dr/dt$ represents
the radial propagation speed of the radio-emitting source.

Therefore, shocks propagating at similar speeds may exhibit
different frequency-drift rates depending on the density
structures encountered along their propagation paths.
This effect may be particularly important for CME--Streamer
interactions, where shocks encounter enhanced coronal
densities and strong density gradients, and for CME--CME
interactions, where preceding eruptions can modify the
upstream plasma conditions and shock geometry.
Furthermore, type II radio emission may originate from
localized regions of a CME-driven shock front, whereas
the LASCO linear speed describes the projected motion
of the CME structure. Consequently, the radio-emitting
shock speed and the measured CME linear speed need not
be identical. This comparison indicates that CME
linear speed alone is insufficient to characterize the
frequency evolution of successive type II radio bursts.
The results are consistent with a possible role for
variations in the coronal density environment and shock
geometry in determining the observed radio signatures.
However, a quantitative comparison between radio-derived
shock speeds and CME speeds would require direct
measurements of individual spectral-lane drift rates
together with an appropriate coronal electron-density model.

\subsection{Implications of the Exploratory Classification Analysis}
\label{sec:section4.3}

The exploratory classification analysis demonstrates that CME linear speed and angular width contain useful, but incomplete, information regarding the physical mechanisms responsible for successive type II radio bursts. The modest classification performance obtained using these two parameters reflects the substantial overlap between the different interaction classes identified in the statistical analyses (see Figure \ref{fig:figure4}. In particular, the overlap between the CME--Streamer and Single CME populations indicates that similar CME kinematics may arise from fundamentally different physical interaction scenarios. This result emphasizes that the occurrence of successive type II radio bursts cannot be explained solely by global CME kinematic parameters. Instead, the production of repeated type II emission is likely controlled by a combination of factors, including the evolution of the CME-driven shock, the magnetic topology of the surrounding corona, the distribution of coronal streamers, local plasma density, and the relative geometry between the propagating shock and ambient coronal structures. These parameters are not captured by the CME speed and angular width alone, explaining the limited predictive capability of the exploratory classification analysis. This is evidence that successive type II radio bursts represent a complex physical phenomenon governed by multiple interacting processes. Incorporating additional observational constraints, including radio imaging, extreme ultraviolet observations, shock kinematics, and coronal magnetic field models, is expected to improve future attempts to identify the physical mechanisms responsible for successive type II emission.
In particular, the overlap between the CME--Streamer and Single CME populations indicates that similar CME kinematics can be generated by fundamentally different physical interaction scenarios. 

\subsection{CME--Streamer and Single CME Events similarity}
\label{sec:section4.4}

We found no statistically significant differences in CME linear speed or angular width are found between the CME-streamer and Single CME populations in the present sample. Their considerable overlap indicates that global CME kinematics alone cannot establish a connection of successive type II radio bursts with an identifiable streamer encounter. This is physically plausible since the interaction mechanism depends not only on the properties of the erupting CME, but also on the structure of the surrounding corona. The Kruskal--Wallis and Dunn's post-hoc tests indicate that the distributions of these parameters substantially overlap, while the multiclass logistic regression and Random Forest classifiers were unable to reliably separate these two populations. This statistical similarity is reflected in the relatively poor predictive performance of all three-class classification models. The observed overlap is physically reasonable because both interaction mechanisms involve only a single eruptive structure propagating through the ambient corona. The location and orientation of the streamer in relation to the expanding CME and its shock are important for a CME-streamer event.
In contrast, a given CME speed and angular width may encounter a streamer in one coronal configuration, but propagate with no clearly identifiable streamer encounter in another. Similarly, repeated type II emission from a single CME may suggest changes in shock conditions as the shock propagates through an inhomogeneous corona even if no external interaction can be identified from the available observations. In CME--Streamer events, the CME-driven shock interacts with a pre-existing coronal streamer, whereas in Single CME events the shock evolves without a clearly identifiable interaction with another large-scale coronal structure. Although these two scenarios differ in terms of the surrounding coronal environment, they may originate from CMEs possessing very similar kinematic properties.

This result suggests that the interaction mechanism cannot be determined solely from the intrinsic properties of the CME. Instead, it depends critically on the large-scale coronal environment through which the CME propagates. Parameters such as the location, orientation, and density structure of nearby streamers, the ambient magnetic field topology, and the propagation direction of the CME are likely to influence whether a CME-driven shock encounters a streamer capable of producing enhanced type II radio emission. The inability of the predictive models to distinguish CME--Streamer from Single CME events therefore does not indicate a failure of the statistical methodology. Rather, it demonstrates that CME speed and angular width alone do not fully capture the physical conditions governing the production of successive type II radio bursts. Incorporating information describing the coronal environment is therefore expected to substantially improve future classification efforts.

\section{Conclusions}
\label{sec:section5}

The occurrence of successive type II bursts should also be considered in the context of the broader CME population. \cite{Kumari2023b} reported that only approximately 4\% and 3\% of CMEs were associated with metric type II radio bursts during solar cycles 23 and 24, respectively, while 48.1\% of the CME-associated type II bursts in their detailed sample were associated with full-halo or partial-halo CMEs. These findings indicate that type II radio emission is associated with only a small fraction of the overall CME population, despite its preferential association with relatively wide and fast eruptions.

The present study addresses a more specific subset of these events, successive type II bursts associated with CME–CME interactions, CME–streamer interactions, and single CMEs. Although the broader CME and type II catalogues provide useful statistical context, the determination of interaction-specific occurrence rates requires a control sample in which CME–CME and CME–streamer interactions have been independently identified, including events without successive type II emission. Such a classification is beyond the scope of the present analysis. Consequently, the differences in CME speed and angular width reported here should be interpreted as conditional properties of the successive type II event population, rather than as direct measures of the probability that a given CME interaction produces successive type II emission.

In the present work, a systematic statistical study of consecutive type II solar radio bursts observed during Solar Cycles 23 and 24 as well as the ascending phase of Solar Cycle 25 has been performed. Using radio dynamic spectra and SOHO/LASCO coronagraph observations, a catalogue of 73 independent successive type II events was constructed. Each event was studied independently and classified according to the inferred physical mechanism responsible for the recurrent type II emission. The primary findings of this study are given below:

\begin{enumerate} 
\item A total of 73 successive type II radio burst events were identified, of which 56 could be unambiguously assigned to one of three physical classes: CME–Streamer interaction, CME–CME interaction or Single CME. The remaining events were classified as \emph{Unclear} or \emph{Not Visible} due to lack of observational evidence.

\item CME--streamer interactions constitute the largest group among the 56 physically classified events, accounting for 50\% of this sample. This predominance highlights the potential role of interactions
between CME-driven shocks and large-scale coronal structures in
successive type II radio emission. However, the relative abundance
of this class in the present catalogue does not establish its
intrinsic probability of producing successive bursts.

\item The CME linear speed shows only small differences for the three interaction classes. CME--Streamer events and Single CME events are generally associated with faster CMEs than CME--CME interaction events, but the differences are not statistically significant when the full classified sample is considered. \item CME angular width has significantly more discriminating power than CME linear speed. The preferential association of wider CMEs with CME–Streamer interaction events implies that the spatial extent of the CME-driven shock is more tightly correlated with the production of successive type II radio bursts than the bulk propagation speed alone.

\item Our results suggest that more physical parameters are needed to distinguish the different interaction scenarios, such as shock evolution, coronal magnetic topology, geometry of the streamers, and the ambient plasma environment.

\item The results presented support where successive type II radio bursts are produced by multiple physical pathways, as opposed to a single universal mechanism. Future studies that combine radio imaging, coronagraph observations, EUV imaging and multi spacecraft measurements will be needed to develop a more complete picture of the physical conditions responsible for repeated shock-associated radio emission.
\end{enumerate}

\begin{acknowledgments}
AK acknowledges the ANRF Prime Minister Early Career Research Grant (PM ECRG) program. We all acknowledge data from National Oceanic and Atmospheric Administration - Space Weather Prediction Center (NOAA SWPC) and CDAW website. We all acknowledge the the OVRO-LWA data. The OVRO--LWA expansion project was supported by NSF under grant AST-1828784. OVRO--LWA solar operations are supported by NSF grant AGS-2436999 to NJIT.
\end{acknowledgments}

\begin{contribution}
AK conceptualized the study, performed the major analysis, and wrote the manuscript. DP contributed to scientific discussions. JK prepared the initial dataset used for the present study.
\end{contribution}

\facilities{SOHO (LASCO), e-Callisto, NOAA/SWPC, CDAW}
\software{SciPy \citep{virtanen2020scipy}, Matplotlib \citep{hunter2007matplotlib}, Pandas \citep{mckinney2010pandas}, Scikit-learn \citep{pedregosa2011scikit}, Statsmodels \citep{seabold2010statsmodels}, Scikit-posthocs \citep{terpilowski2019scikitposthocs}}

\appendix

\section{Catalogue of Successive Type II Radio Bursts}
\begin{longtable}{cllllllllcl}

\caption{Catalogue of the successive type II radio burst events
analysed in this study. The CME type and speed correspond to
the first associated CME.}
\label{tab:successive_typeII_catalogue}\\

\toprule
\multirow{3}{*}{No.} &
\multirow{3}{*}{Date} &
\multicolumn{3}{c}{First Type II Burst} &
\multicolumn{3}{c}{Second Type II Burst} &
\multirow{3}{*}{CME Type} &
\multirow{3}{*}{\shortstack{First CME\\Speed\\(km s$^{-1}$)}} &
\multirow{3}{*}{Interaction Type} \\

\cmidrule(lr){3-5}
\cmidrule(lr){6-8}

&
&
Start &
End &
Frequency &
Start &
End &
Frequency &
&
&
\\

\cmidrule(lr){3-5}
\cmidrule(lr){6-8}

&
&
(UT) &
(UT) &
(MHz) &
(UT) &
(UT) &
(MHz) &
&
&
\\

\midrule
\endfirsthead

\multicolumn{11}{c}{
\tablename\ \thetable\ -- Continued from previous page
}\\

\toprule
\multirow{3}{*}{No.} &
\multirow{3}{*}{Date} &
\multicolumn{3}{c}{First Type II Burst} &
\multicolumn{3}{c}{Second Type II Burst} &
\multirow{3}{*}{CME Type} &
\multirow{3}{*}{\shortstack{First CME\\Speed\\(km s$^{-1}$)}} &
\multirow{3}{*}{Interaction Type} \\

\cmidrule(lr){3-5}
\cmidrule(lr){6-8}

&
&
Start &
End &
Frequency &
Start &
End &
Frequency &
&
&
\\

\cmidrule(lr){3-5}
\cmidrule(lr){6-8}

&
&
(UT) &
(UT) &
(MHz) &
(UT) &
(UT) &
(MHz) &
&
&
\\

\midrule
\endhead

\bottomrule
\endlastfoot

1 & 1997-11-03 & 09:13 & 09:31 & 35--80 & 10:29 & 10:47 & 36--80 & Normal & 338 & CME-Streamer\\
2 & 1998-03-02 & 07:00 & 07:12 & 30--160 & 07:19 & 07:28 & 35--50 & NA & NA & Not Visible \\
3 & 1998-09-28 & 06:55 & 07:02 & 50--200 & 07:05 & 07:12 & 40--80 & NA & NA & Unclear \\
4 & 1998-12-23 & 06:15 & 06:32 & 30--80 & 06:21 & 06:31 & 18--50 & NA & NA & Unclear \\
5 & 1999-01-17 & 19:58 & 20:06 & 25--70 & 19:59 & 20:03 & 30--80 & NA & NA  & Not Visible \\
6 & 1999-06-23 & 05:45 & 05:52 & 33--88 & 05:46 & 05:50 & 25--45 & Normal & 613 & Single CME \\
7 & 1999-07-12 & 18:40 & 18:51 & 30--60 & 19:48 & 20:03 & 30--08 & NA & NA & Not Visible \\
8 & 1999-08-02 & 05:40 & 05:50 & 30--80 & 05:41 & 05:46 & 40--70 & NA & NA & Unclear\\
9 & 1999-08-02 & 05:40 & 05:50 & 30--80 & 06:32 & 06:33 & 80--100 & Partial Halo & 289 & CME-Streamer\\
10 & 1999-08-02 & 05:40 & 05:50 & 30--80 & 06:49 & 06:53 & 30--45 & Partial Halo & 289 & CME-Streamer\\
11 & 1999-08-02 & 05:41 & 05:46 & 40--07 & 06:32 & 06:33 & 80--100 & NA & NA & Unclear \\
12 & 1999-08-02 & 06:32 & 06:33 & 80--100 & 06:49 & 06:53 & 30--45 & Partial Halo & 289 & CME-Streamer \\
13 & 2000-02-17 & 20:25 & 20:40 & 26--180 & 20:38 & 20:49 & 40--150 & Halo & 728 & Single CME \\
14 & 2000-03-18 & 21:03 & 21:16 & 35--113 & 21:56 & 22:01 & 53--190 & Partial Halo &  792& Single CME \\
15 & 2000-07-21 & 14:40 & 14:44 & 72--180 & 14:48 & 14:57 & 30--75 & NA & NA & Unclear \\
16 & 2000-09-01 & 18:27 & 18:36 & 30--75 & 18:47 & 18:52 & 30--52 & Normal & 411 & CME-Streamer \\
17 & 2000-09-12 & 11:43 & 11:52 & 25--48 & 12:07 & 12:13 & 25--45 & Halo & 1550 & CME-CME \\
18 & 2000-09-19 & 08:18 & 08:19 & 25--180 & 08:24 & 08:24 & 77--100 & Normal & 766 & CME-Streamer \\
19 & 2006-12-13 & 02:26 & 02:44 & 25--180 & 02:27 & 02:35 & 30--300 & Halo & 1787 & CME-Streamer \\
20 & 2006-12-14 & 22:09 & 22:13 & 25--180 & 22:10 & 22:13 & 50--140 & NA & NA & Not Visible \\
21 & 2011-02-28 & 07:40 & 07:44 & 57--110 & 07:41 & 07:44 & 59--180 & NA & NA & Unclear \\
22 & 2011-08-02 & 06:08 & 06:22 & 23--90 & 06:17 & 06:28 & 18--57 & Halo & 1042 & CME-CME \\
23 & 2011-09-04 & 04:42 & 04:49 & 57--130 & 04:48 & 04:54 & 45--95 & Normal & 262 & CME-CME \\
24 & 2011-09-06 & 22:19 & 22:34 & 40--300 & 22:36 & 22:38 & 30--45 & Normal & 262 & CME-Streamer \\
25 & 2011-09-20 & 07:00 & 07:12 & 27--60 & 07:04 & 07:11 & 42--31 & Normal & 343 & CME-CME \\
26 & 2011-10-01 & 09:06 & 09:22 & 25--180 & 09:55 & 10:07 & 25--180 & Partial Halo & 448 & CME-Streamer \\
27 & 2011-11-15 & 00:11 & 00:15 & 62--173 & 00:12 & 00:16 & 100--180 & Normal & 285 & CME-Streamer \\
28 & 2012-03-07 & 00:17 & 00:31 & 25--180 & 01:09 & 01:29 & 25--180 & Halo & 2684 & Single CME \\
29 & 2012-06-03 & 17:53 & 18:17 & 25--150 & 17:59 & 18:17 & 25--160 & Halo & 1825 & CME-Streamer \\
30 & 2012-07-31 & 00:00 & 00:14 & 25--87 & 00:19 & 00:29 & 20--50 & NA & NA & Unclear \\
31 & 2013-08-15 & 22:17 & 22:37 & 25--78 & 22:23 & 22:27 & 20--35 & Normal & 359 & CME-Streamer \\
32 & 2013-10-22 & 21:22 & 21:25 & 25--180 & 21:29 & 21:38 & 25--39 & Normal & 589 & CME-CME \\
33 & 2013-10-25 & 07:59 & 08:10 & 25--180 & 08:59 & 09:03 & 45--145 & Halo & 587 & CME-Streamer \\
34 & 2013-10-28 & 04:37 & 04:41 & 25--117 & 04:52 & 04:57 & 35--57 & NA & NA & Unclear \\
35 & 2013-11-05 & 22:13 & 22:22 & 40--450 & 22:20 & 22:27 & 60--180 & Normal & 209 & Single CME \\
36 & 2014-02-10 & 05:15 & 05:18 & 35--57 & 05:22 & 05:24 & 35--50 & Normal & 215 & Single CME \\
37 & 2014-03-06 & 09:29 & 09:29 & 25--180 & 09:39 & 09:39 & 25--180 & Normal & 316 & CME-CME \\
38 & 2014-05-09 & 02:21 & 02:33 & 56--180 & 02:37 & 02:46 & 18--57 & Halo & 1099 & CME-CME \\
39 & 2014-06-01 & 01:42 & 01:46 & 50--90 & 01:50 & 01:52 & 23--40 & NA & NA & Unclear \\
40 & 2014-06-12 & 22:00 & 22:04 & 25--57 & 22:13 & 22:23 & 20--40 & Normal & 970 & CME-CME \\
41 & 2014-08-05 & 15:48 & 16:01 & 25--84 & 15:54 & 15:57 & 31--50 & Normal & 490 & CME-Streamer \\
42 & 2014-09-18 & 08:39 & 08:44 & 42--180 & 08:49 & 08:54 & 36--110 & Normal & 285 & CME-CME \\
43 & 2014-10-02 & 18:58 & 19:11 & 25--180 & 19:31 & 19:38 & 27--45 & Partial Halo & 512  & CME-CME \\
44 & 2014-10-21 & 12:47 & 12:54 & 25--37 & 13:35 & 13:54 & 60--180 & Normal & 260 & CME-CME \\
45 & 2014-10-21 & 12:47 & 12:54 & 25--37 & 13:42 & 13:50 & 27--169 & Normal & 260 & CME-CME \\
46 & 2014-10-21 & 13:35 & 13:54 & 60--180 & 13:42 & 13:50 & 27--169 & Normal & 260 & CME-CME \\
47 & 2014-12-30 & 00:47 & 00:56 & 106--180 & 01:00 & 01:00 & 125--180 & Normal & 873 &Single CME \\
48 & 2015-02-09 & 23:14 & 23:26 & 27--95 & 23:23 & 23:29 & 33--61 & Halo & 1106 & CME-Streamer \\
49 & 2015-10-29 & 02:19 & 02:30 & 25--180 & 02:38 & 02:46 & 30--57 & Partial halo & 530 & CME-CME \\
50 & 2017-09-12 & 07:29 & 07:49 & 25--80 & 07:47 & 07:49 & 25--111 & Normal & 252 & CME-Streamer \\
51 & 2021-08-28 & 05:09 & 05:11 & 30--82 & 06:04 & 06:13 & 41--84 & Partial halo & 191 & CME-Streamer \\
52 & 2022-03-31 & 18:34 & 18:47 & 25--180 & 19:29 & 19:36 & 25--180 & Normal & 641 & Single CME \\
53 & 2022-05-24 & 22:23 & 22:28 & 42--55 & 22:43 & 22:45 & 31--48 & Normal & 569 & CME-Streamer \\
54 & 2022-08-18 & 10:49 & 11:15 & 25--83 & 12:01 & 12:15 & 25--151 & Halo & 1327 & CME-Streamer \\
55 & 2022-09-23 & 13:50 & 14:06 & 25--170 & 14:01 & 14:05 & 25--37 & Partial Halo & 685 & CME-CME \\
56 & 2023-05-05 & 07:06 & 07:19 & 30--122 & 07:54 & 08:04 & 25--61 & Halo & 770 & CME-CME \\
57 & 2023-06-08 & 21:14 & 21:23 & 39--70 & 21:15 & 21:17 & 52--165 & Normal & 451 & CME-CME \\
58 & 2023-07-03 & 21:09 & 21:19 & 31--180 & 21:15 & 21:22 & 30--174 & Normal & 208  & CME-Streamer \\
59 & 2023-07-03 & 21:09 & 21:19 & 31--180 & 21:31 & 21:36 & 25--180 & Normal & 208 & CME-Streamer \\
60 & 2023-07-03 & 21:15 & 21:22 & 30--174 & 21:31 & 21:36 & 25--180 & Normal & 208 & CME-Streamer \\
61 & 2023-07-15 & 10:00 & 10:06 & 34--83 & 10:18 & 10:32 & 41--73 & Normal & 887 & Single CME \\
62 & 2023-07-24 & 17:35 & 18:06 & 32--84 & 17:38 & 18:06 & 25--147 & Halo & 1204 & CME-Streamer \\
63 & 2023-11-28 & 19:34 & 19:44 & 25--180 & 19:35 & 20:03 & 25--180 & Halo & 741 & CME-Streamer \\
64 & 2023-11-28 & 19:34 & 19:44 & 25--180 & 19:59 & 20:07 & 35--47 & NA & NA & Unclear \\
65 & 2023-11-28 & 19:35 & 20:03 & 25--180 & 19:59 & 20:07 & 35--47 & Halo & 741 & CME-Streamer \\
66 & 2024-05-14 & 16:50 & 17:15 & 25--52 & 17:30 & 18:00 & 25--180 & Halo & 2010 & CME-CME \\
67 & 2024-07-17 & 06:39 & 06:45 & 76--134 & 07:00 & 07:06 & 39--58 &  Normal & 220 & CME-Streamer \\
68 & 2024-07-24 & 17:26 & 17:37 & 25--152 & 18:16 & 18:24 & 25--72 & Normal & 580 &  Single CME \\
69 & 2024-07-27 & 04:33 & 04:51 & 75--180 & 05:46 & 05:56 & 30--124 & NA & NA & Unclear \\
70 & 2024-08-02 & 04:49 & 04:59 & 25--138 & 04:49 & 04:59 & 25--131 & Partial Halo & 1141 & CME-Streamer \\
71 & 2024-08-08 & 19:29 & 19:35 & 62--180 & 20:05 & 20:15 & 25--49 & Halo & 789 & CME-Streamer \\
72 & 2024-10-15 & 18:11 & 18:19 & 25--148 & 18:32 & 18:37 & 111--180 & NA & NA & Unclear \\
73 & 2024-10-18 & 23:29 & 23:33 & 109--180 & 23:30 & 23:32 & 71--180 & NA & NA & Unclear \\

\end{longtable}

\section{Event Selection and Classification Criteria}

\label{appendix:B}

The catalogue of successive type II radio burst events presented in this work was developed by means of a two-stage procedure. Candidate successive type II radio burst events were first identified based only on temporal properties of the radio bursts listed in the NOAA/Space Weather Prediction Center (SWPC) Solar Event Reports . Then each candidate event was investigated through radio dynamic spectra and SOHO/LASCO white-light coronagraph observations, so as to derive the corresponding CME configuration and characterise the underlying physical interaction mechanism.

\subsection{Identification of Successive Type II Radio Bursts}

The original event list was prepared from the full NOAA/SWPC catalogue of type II radio bursts from January 1996 to August 2025. Candidate successive events for each type II burst were identified by the reported start and end times. A pair of type II radio bursts was regarded as a \emph{successive type II event} if the start time of the second burst was less than one hour after the end of the first burst. The selection criterion is given mathematically as follows:

\begin{equation}
0 \le
t_{\rm start,2}
-
t_{\rm end,1}
\le
60~{\rm min},
\end{equation}

where $t_{\rm end,1}$ is the end time of the first type II burst, and $t_{\rm start,2}$ is the start time of the second type II burst. This criterion was selected to maintain the temporal continuity of subsequent radio bursts and to avoid considering unrelated solar eruptive events separated by longer time intervals. All candidate pairs fulfilling this condition were then manually checked against the corresponding radio dynamic spectra to ensure that the two emissions were indeed successive type II bursts.

\subsection{Association with Coronal Mass Ejections}

Each candidate successive type II event was studied individually using SOHO/LASCO C2 and C3 coronagraph observations, as well as the SOHO/LASCO CME catalogue. The associated CME(s) were identified by a combination of temporal correspondence, propagation direction, position angle and morphological evolution observed in the coronagraph images. For the multiple-CME events, the relative timing and outward propagation of each eruption were examined to see whether any physical interaction occurred within the LASCO field of view. The radio dynamic spectra were analysed at the same time to determine the temporal relationship between the successive type II bursts and the corresponding CME evolution.

\subsection{Classification of the Physical Interaction Mechanism}

Using the combined radio and coronagraph observations, each successive type II event is placed into one of five physical categories based on the observational criteria shown in Table~\ref{tab:classification}. The classification was performed by visual inspection of all observations available and is the most probable physical explanation for the event.

\begin{table*}[ht]
\centering
\caption{Observational criteria adopted for the classification of successive Type II radio burst events.}
\label{tab:classification}

\begin{tabular}{ll}
\toprule
\textbf{Interaction Class} &
\textbf{Observational Criteria} \\
\midrule

Single CME &
A single CME is temporally associated with the successive Type II radio bursts, with no observational \\ 
& evidence of interaction with another CME or with a coronal streamer during its outward propagation. \\

CME--Streamer &
The CME-driven shock exhibits clear signatures of interaction with a coronal streamer, including \\ 
& streamer deflection, compression, brightening, or other characteristic signatures visible in the coronagraph \\ 
& observations. \\

CME--CME &
Two or more CMEs propagate along similar position angles with temporal overlap and exhibit \\ 
& clear evidence of interaction or overtaking within the LASCO field of view. \\

Unclear &
The available observations do not provide sufficient evidence to uniquely determine the physical \\ 
& mechanism responsible for the successive Type II radio bursts, or multiple interpretations remain equally \\ 
& plausible. \\

Not Visible &
No reliable CME or interaction signature can be identified because of observational limitations, \\ 
& data gaps, poor image quality, or the absence of a clearly detectable CME. \\

\bottomrule
\end{tabular}
\end{table*}

All five categories are included in the full catalogue to preserve the complete observational record. However, the \emph{Unclear} and \emph{Not Visible} classes were removed from the statistical analyses presented in Sections~3.3--3.6 as the underlying physical interaction mechanisms could not be established with sufficient confidence. Therefore, all the quantative analysis was performed using only three physically intelligible categories: CME--CME, CME--Streamer, and Single CME.

\section{Statistical and Exploratory Classification Methods}
\label{appendix:C}

This appendix summarizes the statistical procedures used to compare the physical classes of successive Type II radio burst events and the exploratory classification analysis performed using CME kinematic parameters. All statistical analyses were carried out using Python with the \texttt{NumPy}, \texttt{Pandas}, \texttt{SciPy}, \texttt{scikit-posthocs}, \texttt{scikit-learn}, and \texttt{Matplotlib} libraries.

\subsection{Descriptive Statistics}

Separate descriptive statistics were computed for the CME linear speed and angular width for each interaction class (CME–CME, CME–Streamer, and Single CME). These were sample size, mean, median, standard deviation, minimum, maximum and interquartile range. The resulting statistics are summarised in Table~\ref{tab:statistics}.

\subsection{Non-parametric Statistical Tests}

Due to the relatively small sample sizes and the non-Gaussian distributions of the CME parameters, non-parametric statistical methods were adopted throughout this study. The differences between the three classes of interaction were initially evaluated with the Kruskal--Wallis rank-sum test. This test tests the null hypothesis that the three independent samples are drawn from the same distribution, but without any assumption of normality. The Kruskal--Wallis test was used to verify the existence of a statistically significant difference ($p<0.05$) between the classes of interaction and pairwise comparisons between the classes were performed with Dunn's multiple-comparison test with Holm correction to take into account multiple testing.

\subsection{Effect Size Estimation}

Statistical significance does not provide a metric of the size of the observed differences. Cliff’s Delta ($\delta$), a non-parametric effect size measure, was computed for all pairwise comparisons. The generally accepted interpretation is that

\begin{center}
\begin{tabular}{lc}
\toprule
$|\delta|$ & Interpretation\\
\midrule
$<0.15$ & Negligible\\
$0.15-0.33$ & Small\\
$0.33-0.47$ & Medium\\
$>0.47$ & Large\\
\bottomrule
\end{tabular}
\end{center}

\subsection{Exploratory Classification Analysis}

To verify whether global CME kinematic properties provide predictive information about the physical interaction mechanism, we constructed an exploratory multinomial logistic regression classifier using CME linear speed and angular width as predictor variables. The analysis only considered the 56 physically classified events of the CME--CME, CME--Streamer and Single CME categories. This was done in an effort to reduce the variance of a single train-test split, while still maintaining the proportions of the classes within each fold. Model performance was assessed using repeated stratified five-fold cross-validation.

The reported classification accuracy, macro-averaged precision, recall and F$_1$-score correspond to the mean and standard deviation over all cross-validation realisations. The classification experiment is meant to be only an exploratory evaluation of the discriminatory information embedded in the CME parameters and should not be considered as a predictive operational model, given the small sample size.

\begin{table}[ht]
\centering
\caption{Summary of the statistical methods used in this work.}
\label{tab:statmethods}
\begin{tabular}{ll}
\toprule
Analysis & Method \\
\midrule
Descriptive statistics & Mean, median, quartiles, standard deviation \\
Group comparison & Kruskal--Wallis test \\
Pairwise comparison & Dunn's test (Holm corrected) \\
Effect size & Cliff's Delta \\
Classification & Multinomial logistic regression \\
Validation & Repeated stratified 5-fold cross-validation \\
\bottomrule
\end{tabular}
\end{table}

\bibliography{references}{}
\bibliographystyle{aasjournalv7}

\end{document}